\documentclass{emulateapj}
\def    \mc {\mbox{ cos }}
\def    \ms {\mbox{ sin }}

\def \bea {\begin{eqnarray}}
\def \ena {\end{eqnarray}}                  
\def \bee {\begin{equation}}
\def \ene {\end{equation}}

\begin{document}
\shorttitle{Superparamagnetic Relaxation}
\shortauthors{Lazarian \& Hoang}
\title{ Alignment of Dust with Magnetic Inclusions: Radiative Torques and Superparamagnetic Barnett and Nuclear Relaxation}
\author{A. Lazarian \& Thiem Hoang}
\affil{Dept. of Astronomy, University of Wisconsin,
   Madison, WI53706; lazarian; hoang@astro.wisc.edu}
\begin{abstract}

We consider grains with superparamagnetic inclusions and report two new condensed matter effects 
that can enhance the internal relaxation of the energy of a wobbling grain, namely, superparamagnetic Barnett relaxation,
as well as, an increase of frequencies for which nuclear relaxation becomes important. This findings
extends the range of grain sizes for which grains are thermally trapped, i.e. rotate thermally, in spite of the presence
of uncompensated pinwheel torques.
In addition, we show that the alignment of dust grains by radiative torques gets modified for superparamagnetic grains,
with grains obtaining perfect alignment with respect to magnetic fields as soon as the grain gaseous randomization time
gets larger than that of paramagnetic relaxation. 
The same conclusion is valid for the mechanical alignment of helical grains. If observations confirm that the degrees of
alignment are higher than radiative torques can produce alone, this may be a proof of the presence of superparamagentic
inclusions.
\end{abstract}

\section{Introduction}

Alignment of dust in magnetic fields is one of the astrophysical problems of the longest standing. 
Discovered by Hiltner (1949) and Hall (1949) via observation of the polarization of starlight attenuated by 
interstellar dust, aligned grains are shown to trace magnetic fields in various environments, 
from interplanetary space (see Rosenbush et al. 2007) to circumstellar regions 
(see Aitken et al. 2002; Cho \& Lazarian 2005). 

The presently-accepted dominant alignment process is based on the action of radiative torques (RATs) on irregular grains (see
Lazarian 2007 for a review).
This process was proposed by Dolginov \& Mytrophanov (1976) and extended, corrected and elaborated in
the subsequent publications (Lazarian 1995; Draine 1996; Draine \& Weingartner 1996, 1997, henceforth, DW96 and DW97; Weingartner \& Draine 2003,
Weingartner 2006; Lazarain \& Hoang 2007a, henceforth LH07; Hoang \& Lazarian 2008). This has induced theory-based modeling of polarized radiation from molecular clouds (Cho \& Lazarian 2005; Pelkonen et al. 2007,
Bethell et al. 2007) and accretion disks (Cho \& Lazarian 2007). 

While DW96 professed the omnipotence of RATs for aligning interstellar grains, the physics involved happens to be more complex than it was initially anticipated.
In particular, an analytical model (AMO) for RATs proposed in LH07 shows that, for a variety of the directions
between the magnetic field and the radiation flux, the alignment is not perfect (see also a particular example of this in Weingartner \& Draine 2003 and more details
 in Hoang \& Lazarian 2008). 
This causes the justified worries whether all the essential physical processes has been accounted for to explain the high degrees of interstellar polarization
(see Whittet et al. 2008). 

In LH07a we showed that ordinary paramagnetic relaxation does not affect the RAT alignment in most astrophysical environments, e.g., in interstellar gas.
This does not preclude inclusions with strong magnetic response, e.g. superparamagnetic inclusions, to affect the RAT alignment. Such grains were discussed in Jones \& Spitzer (1967) and also Martin (1995). Given the abundance of iron in dust, this would not be a far fetched suggestion. Moreover studies of interstellar grains caught in 
the Earth atmosphere (see Bradley 1994; Goodman \& Whittet 1995), as well as, laboratory studies of the silicate films (see Djouadi et al. 2007) are suggestive of
the abundance of such inclusions.  
Below in \S 2 we consider both the effect of superparamagnetic inclusions on
grain internal relaxation, i.e. on the relaxation that  may induce grain to
rotate about its axes of maximal inertia (Purcell 1979; Lazarian 1994; Lazarian \& Roberge 1997; Lazarian \& Efroimsky 1998; Lazarian \& Draine 1999ab, henceforth LD99ab)  and on the RAT alignment (in \S 3).

\section{Barnett and Nuclear Relaxation in Superparamagnetic Grains}

 A general susceptibility of the grain with superparamagnetic inclusions is given in Draine \& Lazarian (1999). 
If, however, we consider
only low frequency response of the material, superparamagnetic inclusions containing $N_{cl}$ atoms act in the external 
magnetic field as magnetic moments of the moment $N_{cl}\mu$, where $\mu$ is the moment of an individual atom. This 
increases the grain magnetic susceptibility by a
factor of $N_{cl}$, which may vary from approximately $10$ to $10^6$ (see Kneller \& Laborsky 1963; Billas et al. 1994). The enhanced
susceptibility affects the internal relaxation through the Barnett effect. 

Barnett  effect (Barnett 1915) can be well understood classically (see LD99a for a quantum mechanical treatment). If one considers, instead of an electron spin, a gyroscope with a momentum
${\bf J}$. As the grain rotates with angular velocity ${\bf \Omega}$, the torque acting on a spin is  $\frac{d}{dt} {\bf J}={\bf \Omega}\times {\bf J}$. The equivalent torque can be induced by a magnetic field ${\bf H}_{B}$ acting along the
grain angular velocity axis, i.e. $1/c {\bf \mu}\times {\bf H}_{B}$, which provides the Barnett-equivalent field
 $H_{eqv}=\frac{J c}{\mu} \Omega$. The magneto-mechanical ratio for an electron is $\gamma=\mu/(Jc)=e/mc$, 
where $e$ and $m$ are the electron
charge and mass, respectively. Thus the Barnett-equivalent field is $H_{eqv}=\Omega/\gamma$.

The Barnett-equivalent magnetic field induces grain magnetization ${\bf M}=\chi {\bf H}_{eqv}= \chi {\bf \Omega}/\gamma$.
Purcell (1979) noticed that when a grain wobbles, the precession of ${\bf \Omega}$ in grain axes induces the
precession of the vector of magnetization ${\bf M}$, which causes energy dissipation. For a grain
at an absolute zero temperature this would result in the grain getting to rotate
about its axis corresponding to the maximal moment of inertia. For grains at finite temperatures, the relaxation induces
important coupling between rotational and vibrational degrees of freedom, which results in many important effects
 (Lazarian 1994, Lazarian \& Roberge 1997, LD99a,b). We expect for grains with enhanced magnetic 
susceptibilities, i.e. larger $\chi$ both the magnetization and the relaxation that this magnetization entails to be enhanced. 

To quantify the effect, consider the internal relaxation associated with the Barnett magnetization of a 
wobbling oblate grain. The angle
$\theta$ between the axis of major inertia ${\bf a}_{1}$ and the angular momentum ${\bf J}$ changes according to 
LD99a as
\begin{equation}
\frac{d\theta}{dt}=-(h-1) \frac{VJ^2}{I_{\|}I_{\bot}^{2}\gamma^2} (\chi^"/\omega) \sin\theta \cos\theta f(h,\theta),
\label{theta}
\end{equation}
where $I_{\|}$ and $I_{\bot}$ are the moments of inertial parallel and perpendicular to the axis of major inertia, $h\equiv
I_{\|}/I_{\bot}$, $\omega=(h-1)J\cos\theta/I_{\|}$ is the precession frequency, $V$ is the grain volume, $\chi^"$ is the imaginary
part of magnetic susceptibility, and $f(h,\theta)$ is
a function of order unity that takes into account details of spin-lattice coupling. For sufficiently slow rotation, e.g.
$\Omega<10^{6}$ this function is identically unity and Eq.~(\ref{theta}) coincides with the result in Purcell (1979).

If the magnetic moment of electrons on the outer partially-filled shell is $p\mu_B$, where $\mu_B\equiv e\hbar/2m_e c$ is
the Bohr magneton, while for bulk Fe we take $p=2.22$ (Morish 1980), then for zero-frequency susceptibility of a grain
is (see Draine 1996)
\begin{equation}
\chi_{sup} (0)\approx 1.2 \times 10^{-2} N_{cl} f_p \hat{n}_{23} \hat{p}_{3}^2 \hat{T}_{15}^{-1},
\label{chi}
\end{equation}
 where $f_p\approx 0.01$ is the fraction of Fe atoms, $\hat{n}_{23}\equiv/(10^{23}~cm^{-3})$ and $\hat{T}_{15}\equiv T/(15~K)$ are
 normalized grain density and temperature, respectfully. Superparamagnetic clusters undergo  thermally activated remagnetization at a rate
 \begin{equation}
 \tau^{-1}_{sp}\approx \nu_0 \exp\left[-N_{cl}T_{act}/T\right]
 \label{sp}
 \end{equation}
where $\nu_0\approx 10^{9}$ and $T_{act}\approx 0.011K$ (see Morish 1980)
 
 Adopting critically damped susceptibility from 
Draine \& Lazarian (1999) one can write the imaginary part of the grain susceptibility as 
$\chi^{"}=\chi_{sup}(0)\omega \tau/(1+(\omega\tau/2)^2)^2$
.  Combining  the latter expression for $\chi^{"}$ with Eqs.~(\ref{theta}) and (\ref{chi}) one obtains 
for an $a\times a\sqrt{3} \times a \sqrt{3}$ brick the time of alignment equal to 
\begin{equation}
\tau_{Bar,sup}\approx10^{8}\hat{\rho}^{2}a_{-5}^{7}\frac{1}{N_{cl}}\left(\frac{J_{d}}{J}\right)^{2}\times \left(1+\left(\frac{\omega_{1}\tau}{2}\right)^{2}\right)^{2},\label{tbar}
\label{tau}
\end{equation}
which is a factor $\sim N_{cl}$ smaller than the usual Barnett relaxation time. 

The competing process of internal relaxation, namely, ``nuclear relaxation'' was introduced in LD99b. While the magnetization arising from nuclear spins 
of a rotating grain is $\sim 1000$ times
smaller than the magnetization arising from electron spins, the nuclear relaxation $\sim H_{eqv}^2$
is approximately $\sim 10^6$ times faster than the Barnett relaxation in normal paramagnetic material.
This difference in the value of Barnett-equivalent field $H_{eqv}$ can be 
understood if we recall that the flipping of both electron and nuclear spins in order to get spins
mostly parallel to grain angular velocity is entirely mechanical effect. For a given temperature the
equilibrium excess of oriented spins depends only the angular momentum of the spin and the 
temperature (see Lazarian \& Draine 1999a). For both nuclear and electron spins the former values
are equal of $\hbar$. However, to get the same excess of orientation acting by external magnetic
field on much weaker nuclear moments, one has to use a stronger magnetic field.

\begin{figure}
\includegraphics[width=0.5\textwidth]{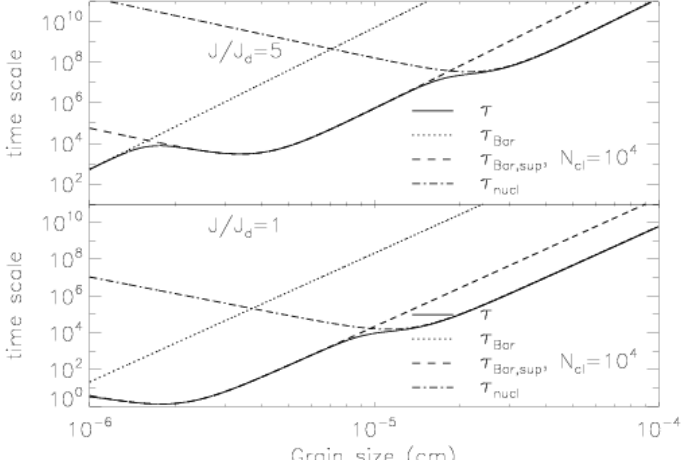}
\caption{Time scales of internal relaxations: Barnett relaxation ($\tau_{Bar}$), Barnett superparamagnetic relaxation ($\tau_{Bar,sup}$), nuclear relaxation ($\tau_{nucl}$) and internal relaxation time ($\tau_{int}$) as functions of grain size, for two values of angular momentum $J/J_{d}=5$ ({\it upper}) and $J/J_{d}=1$ ({\it lower}). Here we adopt number of iron cluster $N_{cl}=10^{4}$. Relaxation by superparamagnetic inclusion is dominant for grains smaller than $3\times 10^{-5}$cm for the case $J/J_{d}=1$. }
\label{f0}
\end{figure}

Superparamagnetism does not change
the amplitude of the nuclear relaxation rate, but, nevertheless, affects the rate of the wobbling at which the relaxation
gets saturated. Indeed, the relaxation of the system of spins depends on the rate of spin precession in the internal varying 
magnetic field of the neighboring magnetic moments. In the medium with magnetic moments $\mu_m$ with
density $n_m$ one can use 
van Vleck (1937) expression for the internal magnetic field $H_i\approx 3.8 n_m \mu_m$. If the medium is extended, $H_i$
does not depend on whether the magnetic moments are clustered or not, as $n_m\sim N_{cl}^{-1}$, while $\mu_m\sim N_{cl}$.
If $\tau_{sp}^{-1}$ is much faster than the  time of nuclear moments precession $\omega_N\approx g_N \mu_N H_i/\hbar$, where $\mu_N\equiv e\hbar/m_p c$ 
is nuclear magneton,
then the fluctuating magnetic field will induce a random walk with the step $\sim \tau_{sp}$. The change in direction of the nuclear moment per step is
$\delta \theta \approx \omega_N \tau_{sp}$, which means that the initial phase relationship is lost after a time 
\begin{equation}
\tau_{nsp}\approx (1/\delta \theta)^2 \tau_{sp}\approx \omega_N^{-1}/(\omega_N \tau_{sp}),
\label{nsp}
\end{equation} 
which should be compared with electron-nuclei decorrelation time $\tau_{ne}\approx \omega_N^{-1}(\omega_{e}/\omega_N)$ and a comparable rate of nuclei-nuclei 
decorrelation $\tau_{nn}\approx 0.58 \tau_{ne} (n_e/n_n)$, where $n_e$ and $n_n$ are densities of unpaired electron and nuclear spins, respectively (LD99b). Therefore,
if the rate of the nuclear relaxation $\tau_{nsp}^{-1}$ can be increased in the presence of superparamagnetic inclusions by a factor $\tau_{sp} \omega_e$, which 
scales according to eq.~(\ref{sp}) as $\mbox{exp}\left(N_{cl} T_{act}/T\right) \omega_e$, provided that, nevertheless $\tau_{sp}^{-1}\gg \omega_N$. This means that for clusters with $N_{cl}<1.5 \times 10^{4}$ the extension of the range of frequencies
for the dominant nuclear relaxation exist. If clusters are larger, then superparamagnetically enhanced nuclear relaxation gets suppresses by the 
factor $(1+(\omega \tau_{nsp}/2)^2)^2$ similar to eq.~(\ref{tau}). Figure \ref{f0} illustrates the relative roles of the Barnett and nuclear relaxation processes in paramagnetic and superparamagnetic grains and shows that for $N_{cl}= 10^4$, nuclear relaxation get modified in the range of grain sizes smaller than $2\times 10^{-5}$ cm. As a result, for grains in the range from $3\times 10^{-6}$ to $2\times 10^{-5}$ cm superparamagnetic Barnett relaxation dominates.

The actual relaxation rate is important for getting the rates of grain flipping.  The corresponding expression according to LD99a is
$\frac{t_{tf}}{\tau_{int}}=\mbox{exp}\left(J^{2}/J_{d}^{2}-0.5\right)$ 
where $\tau_{int}^{-1}$ is the internal relaxation rate, which is the sum of the internal relaxation rates acting on the grains,  $J$ is the magnitude of the angular momentum, and $J_{d}=\sqrt{2I_{1}k_{B} T_{d}}$ is the angular momentum of dust grain corresponding to the dust temperature $T_{d}$; $I_{1}$ is the moment of the axis of major inertia $\bf{a}_{1}$ and $k_{B}$ is the Boltzmann constant.

\section{RAT alignment of superparamagnetic grains}

\begin{figure}
\includegraphics[width=0.5\textwidth]{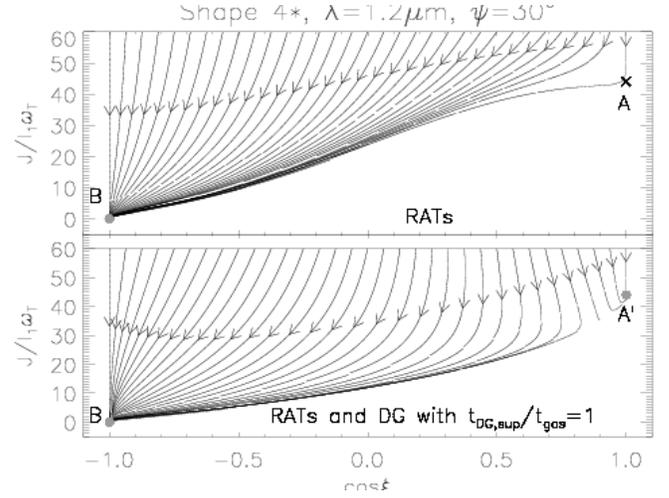}
\caption{Trajectory maps for the alignment by RATs only ({\it upper panel}) and by RATs
  plus superparamagnetic relaxation with $t_{DG,sup}/t_{gas}=1$ ({\it lower
    panel}) for Shape 4* and one light direction $\psi=30^{\circ}$. Unlike in LH07, only trajectories corresponding to ${\bf J}$ parallel to grain axis ${\bf a}$ are shown.
    Repellor points A in the upper panel becomes the attractor point A' in the lower panel
  as a result of superparamagnetic alignment.}
\label{f1}
\end{figure}

 As we see in Figure \ref{f1} the DG relaxation within superparamagnetic material
can influence grain dynamics. The maps shown follow the evolution of an irregular grain (Shape 4* in LD07, i.e., modified from Shape 4 by changing the helicity from left to right), keeping track of
its angular momentum $J$ and the angle $\xi$ between ${\bf J}$ and the magnetic field.
For simplicity, we disregard thermal wobbling and flipping (see Hoang \& Lazarian 2008 for more details).
Comparing the panels of Figure \ref{f1}, we see that without the DG relaxation (or in its presence, but
assuming that grains are paramagnetic), only lower attractor point B  is being observed in the phase
trajectory map, while stationary point A is a repellor point (upper panel). However, this repellor point A becomes a new high attractor point A' when the grain is 
superparamagnetic (lower panel).  

When gaseous bombardment is included, then, as shown in Hoang \& Lazarian (2008), grains are being
kicked out from the low-$J$ attractor point and end up at the high-$J$ attractor point.  Other randomization
processes, e.g. shaking of a flipping grain by ${\bf B}\times {\bf v}$ electric field arising from grain drift (Weingartner 2006) or  from
grain emission, plasma-grain interactions etc. (see Draine \& Lazarian 1998) will
have a similar effect. As a result, grain alignment with
$R\approx 1$, i.e. perfect alignment of grains in respect to magnetic field, is achievable.

What are the degrees of the paramagnetic enhancement that are required to create high-$J$ attractor
points? The answer depends on the angle between radiation direction and the magnetic field $\psi$ and grain properties, that, according to LH07a, 
can be conveniently parameterized by the ratio $q^{max}=Q_{e1}^{max}/Q_{e2}^{max}$,
where $Q_{e1}^{max}$ is the amplitude of the component of the torques parallel to vector ${\bf k}$ and $Q_{e2}^{max}$ is the amplitude of the
component of the torques in the plane of grain axis of major inertia ${\bf a}_{1}$ and ${\bf k}$.
Let us estimate roughly when there are high-$J$ attractor points in the presence of superparamagnetic inclusion. For simplicity, we 
consider the case $\psi=0^{0}$, and the default AMO with the mirror at an angle $\alpha=\pi/4$. Thus the aligning torque reads (see eq. 25 in DW97)
\bea
\langle F(\xi)\rangle_{\phi}&=\frac{M}{J'} \left(-Q_{e1}^{max}(3\mc^{2}\xi-2)\ms\xi+Q_{e2}^{max}\ms2\xi\mc\xi\right)\nonumber\\
&-\frac{\ms\xi \mc\xi}{t'_{DG,sp}},
\label{dis4}
\ena
where the first term containing a factor $M=\frac{u_{rad}\lambda a_{eff}^{2}t_{gas}}{2 I_{1}\omega_{T}}$, where $u_{rad}$ is the radiation energy density, $\lambda$ is the wavelength,
$a_{eff}$ the effective grain size defined as the radius of the spherical grain of the same volume, $t_{gas}$ is the gaseous damping time,  denotes the aligning component of RATs (see eq. ~38 in LH07) and $J'=J/I_{1}\omega_{T}$, where $I_1$ is the maximal moment of inertia and $\omega_T$ is thermal rotational rate at $T=T_{gas}$. The second term represents the contribution from the super-paramagnetic relaxation with characteristic time
\bea
t'_{DG,sp}=\frac{t_{DG,sp}}{t_{gas}}\approx 10^{2} N_{cl}^{-1}\hat{B}^{-1}\hat{K}^{-1}b_{-5}(\frac{\hat{n}\hat{T}_{g}^{1/2}}{\hat{\rho}}).\label{dis3}
\ena

It is obvious from Eq. (\ref{dis4}) that the position of the stationary points $\xi_{s}=0$ and $\pi$ (i.e. $\langle F\rangle_{\phi}=0$) is not affected by the
presence of the additional relaxation. Also, the magnitude of spin-up torque $\langle H\rangle_{\phi}$ at this stationary point is not affected by the superparamagnetic relaxation either (see eq. 26 in DW97). As a result, the stationary point $\xi_{s}$ is an attractor point when
\bea
\left. \frac{1}{\langle H\rangle_{\phi}}\frac{d\langle F\rangle_{\phi}}{d\xi}\right|_{\xi=\xi_{s}}&<0.\label{cri}
\ena
At the stationary point $\xi_{s}=0$, $\langle H\rangle_{\phi} >0$, therefore, using Eq. (\ref{dis4}) and (\ref{cri}) one gets 
\bea
\frac{M}{J_{s}}\left(-Q_{e1}^{max}+2Q_{e2}^{max}\right)-\frac{1}{t'_{DG,sp}}<0.
\ena
\bea
t'_{DG,sp} < \frac{J_{s}}{M\left(-Q_{e1}^{max}+2Q_{e2}^{max}\right)}=\frac{Q_{e1}^{max}}{\left(-Q_{e1}^{max}+2Q_{e2}^{max}\right)},\label{dis5}
\ena
where we have used $J_{s}/M=\langle H(\xi_{s}=0)\rangle_{\phi}=Q_{e1}^{max}$ is the spin-up torque at the stationary point $\xi_{s}=0$. 
For $Q_{e1}^{max}>2Q_{e2}^{max}$, the criterion (\ref{dis5}) is always fulfilled, i.e. there is high-$J$ attractor points regardless of DG relaxation. This is similar to what obtained in LH07.
When  $Q_{e1}^{max}<2Q_{e2}^{max}$, Eq. (\ref{dis5}) gives $ t'_{DG,sup}<{q^{max}}/({2-q^{max}})$ with
 $q^{max}=Q_{e1}^{max}/Q_{e2}^{max}$.
Thus, the AMO predicts that for $q^{max}=0.8$,  the enhancement of paramagnetic susceptibility necessary to convert the
repellor points into attractor points is $t'_{DG,sup}\le 0.67 $ for $\psi=0^{0}$. The same value $q^{max}$ corresponds to
the Shape 4* grain at $\lambda=1.2\mu m$ and for this grains we get the necessary value that we obtain numerically $t'_{DG,sup}\approx 0.56$, which is close
to what the AMO gives us. For experimentally studied in LH07  grain
shapes the corresponding $q^{max}$ varied from $0.4$ to $10$ and  for $\psi$ in the range from $0^{\circ}$ to $\sim 90^{\circ}$  and found that the minimal value $t'_{DG,sup}\approx 5\times 10^{-2}$ is sufficient to produce the high-J attractor points in all cases. This is easily achievable for 
grains having superparamagnetic inclusions with $N_{cl}\sim 5\times 10^{3}$ iron cluster. It is clear that such superparamagnetic inclusions can substantially enhance the degree of grain alignment.

To distinguish the RAT alignment in which magnetic field is only important in terms of inducing
fast Larmor precession, from a hybrid type alignment that includes an additional process, namely, superparamagnetic relaxation 
we shall call the latter process superamagnetic RAT alignment, or, in short, {\it SRAT alignment}.  Note that  superparamagnetic inclusions
also enhance the rate of Larmor precession of ${\bf J}$ in respect to ${\bf B}$, which makes the alignment
with respect to ${\bf B}$ preferable to the radiation vector ${\bf k}$ (see discussion in LH07a). This may allow more reliable interpretation
of polarimetry in terms of magnetic field in circumstellar regions.

\section{Discussion and Summary}

In terms of the RAT alignment it is possible to show that even inclusions with
$N\sim 10^3$ can result in the appearance of a new high-$J$ attractor point. As a result, the RAT
alignment can be enhanced substantially, becoming nearly perfect. This alignment is different
from the alignment described in Draine \& Weingartner (1996), however. In that paper, it was assumed
that RATs provide the spin-up similar to pinwheel torques in Purcell (1979), while ordinary paramagnetic
dissipation does the job of alignment. The latter interpretation of the RAT action is erroneous, as RATs provide
the alignment of their own, making the weak paramagnetic torques irrelevant. It is only after being enhanced
through the presence of superparamagnetic inclusions, that the paramagnetic dissipation can affect the
alignment. If observations prove that perfect alignment of grains is required, {\it SRAT alignment} can achieve this.
Interestingly enough, the synthesis of pinwheel torques and superparamagnetic inclusions proposed in Spitzer \& McGlynn (1979),
does not work for interstellar medium, as superparamagnetic grains are strongly thermally trapped and cannot
rotate suprathermally, unless RATs are present. Note, that  superparamagnetic inclusions will also enhance the mechanical alignment
of helical grains. Such an alignment was proposed in Lazarian \& Hoang (2007b) where it was shown
that  torques arising from the gas-grain drift may act on irregular grains the same way as RATs do.

Our results above can be briefly summarized as follows:

$\bullet$ Superparamagnetic inclusions affect both Barnett and nuclear relaxation, but in a different way:
Barnett relaxation is being enhanced by the increase of magnetic susceptibility, while the nuclear relaxation
does not changes its amplitude, getting, nevertheless extended for higher frequencies.
 
$\bullet$ Supraparamagnetic inclusions modify the RAT alignment and RAT-superparamagnetic mechanism, which we
termed {\it SRAT alignment}, may provide higher degree of alignment compared to the RAT mechanism alone.

$\bullet$ The results above are also applicable to mechanical alignment of irregular grains.

{\bf Acknowledgments} The work was supported by the NSF Center for Magnetic Self-Organization in Laboratory and Astrophysical
Plasmas and NSF grant AST 0507164.

\end{document}